\documentclass[twocolumn]{IEEEtran}

\usepackage[numbers,sort&compress]{natbib}

\usepackage{graphicx}
\usepackage{multicol}
\usepackage{booktabs}
\usepackage{array}
\usepackage{tabularx}
\usepackage[hyphens]{url}
\usepackage[hidelinks]{hyperref}
\hypersetup{breaklinks=true}
\usepackage{array}
\usepackage{hyperref}
\usepackage{balance} 

\usepackage{cite,array,hyperref,balance,tabularx}

\graphicspath{ {./Image/} }

\usepackage{authblk}
\usepackage{tikz}

\usepackage{amsmath}
\usepackage{pifont}

\graphicspath{ {images/} }

\hypersetup{
    colorlinks=false,
    linkcolor=blue,
    filecolor=magenta,
    urlcolor=cyan,}
    
\definecolor{LightCyan}{rgb}{0.88,1,1}

\newcolumntype{a}{>{\columncolor{LightCyan}}c}

\title{Blockchain for Decentralization of Internet: \\Prospects, Trends, and Challenges}

\author[1]{Javad Zarrin
\thanks{$^{1}$
        {\tt\small javad.zarrin@aru.ac.uk}}%
        }
\author[2]{Phang Hao Wen
\thanks{$^{2}$
        {\tt\small edison.phang@student.anglia.ac.uk}}%
        }
\author[3]{Lakshmi Babu-Saheer
\thanks{$^{3}$
        {\tt\small lakshmi.babu-saheer@aru.ac.uk}}%
        }
\author[4]{Bahram Zarrin
\thanks{$^{4}$
        {\tt\small baza@dtu.dk}}%
}

\affil[1-3]{Anglia Ruskin University,
            Cambridge, United Kingdom } 
\affil[4]{Technical University of Denmark (DTU),  Lyngby, Denmark  }

\begin{document}

\maketitle

\begin{abstract}

Blockchain has made an impact on today's technology by revolutionizing the financial industry in its utilization on cryptocurrency and the features it provided on decentralization. With the current trend of pursuing the decentralized Internet, many methods have been proposed to achieve  decentralization considering different aspects of the current Internet model ranging from infrastructure and protocols to services and applications. This paper focuses on using Blockchain to provide a robust and secure decentralized computing system. The paper conducts a literature review on Blockchain-based methods capable for the decentralization of the future Internet. To achieve that decentralization, two research aspects of Blockchain have been investigated that are highly relevant in realizing the decentralized Internet. The first aspect is the consensus algorithms, which are  vital components for decentralization of Blockchain. We have identified three consensus algorithms being PoP, Paxos, and PoAH to be more adequate for reaching consensus in Blockchain-enabled Internet architecture. The second aspect that we investigated is the impact of future Internet technologies on Blockchain, where their combinations with Blockchain would help to make it overcome its established flaws and be more optimized and applicable for Internet decentralization.

\end{abstract}

\section{Introduction}

     \begin{figure*}
         \centering
         \includegraphics[width=140mm]{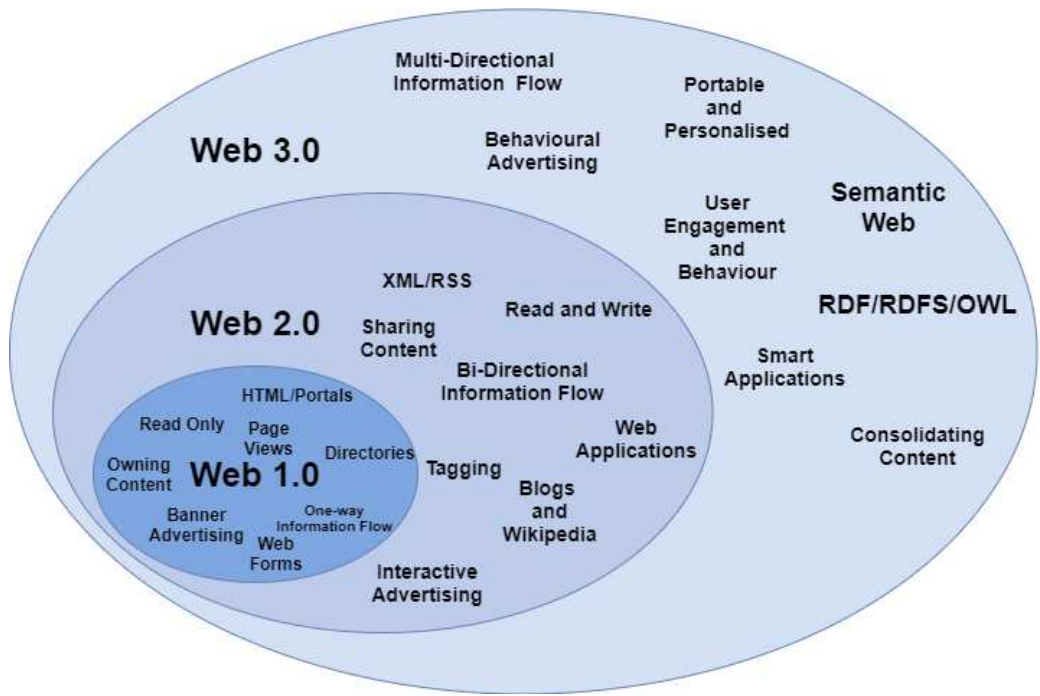}
         \caption{Generations of Web Technology}
         \label{fig:webGeneration}
     \end{figure*}
In the prospect of finding ways to further improve the existing Internet model, there are two ideologies being spearheaded for the development of future Internet. The first solution as suggested by \citep{web3.0Semantic} that seeks to connect every information with Semantic technology to be united into a singularity. The alternate solution is by \citep{web3.0BC} where it aims to decentralize the Internet for equal role power to prevent monopolization from online services. Web 2.0 introduced online services that brought in flaws of requiring centralized services, which is seen in client-server model. To solve this centralized dependency, decentralization would be able to resolve this flaw \citep{mastodon, consortiumDNS, blockDNS}. According to \citep{web3.0BC}, the popularity of decentralization has been attached to Blockchain due to its success in decentralization for cryptocurrencies. 
     
     The centralization of the Internet is not accomplished from a single night, but the gradual development of the Internet and its services over the years. The  introduction of centralized services provided convenience and enabled centralization to be flourished within the Internet. Although these centralized services have provided numerous advancements that have made up the current Internet, the bottom line is that a centralized service would still exhibit a centralized network's vulnerabilities that would jeopardize the network. By having users relying on a centralized service, the users are opened to various types of attacks like Distributed Denial-of-service (DDoS) that could have been easily mitigated through decentralization.

     The main motivation for this work came from the acknowledgment of reliance on centralized systems within the Internet \citep{understandWeb2.0}. It is clear that according to \citep{p2pPrivacy&Decentralization} there has been a push for the development of the Internet to be consolidated into a central overseeing figure for administration. This matter of consolidation with information data is further provoked by privacy concerns caused by large organizations as part of the Big Data scheme.  
     
     Combating this centralization is achievable through decentralization with Blockchain. Blockchain has always been classified as a disruptive technology due to its impact for providing a decentralized solution for communication and transaction. Which brings us to Blockchain's consensus algorithms being capable of enforcing equal roles between peers. This enforcement would also keep these online services in check, preventing centralized power. The aspiration to obtain decentralization is broadened with the trend of implementing Blockchain into Internet Of Things (IoT), and to account for scaling to support the Big Data of the future Internet.
     
     The Internet is a tremendously scaled, geographically distributed, global system of interconnected computer networks that uses the Internet protocol suite (TCP/IP) for communications across nodes and networks. It comprises various components, including infrastructures, hosts, devices, protocols, operating systems, services, and applications.  \\Throughout this paper, we frequently use the terms "decentralized Internet" and "Internet decentralization" to represent the concept of applying possible decentralized approaches in various levels and into any Internet components (e.g., decentralized protocols, applications, and infrastructure) in general and particularly for Web (so-called Decentralized Web, dWeb or Web 3.0). The original Web 1.0 introduced communication with Hypertext Transfer Protocol (HTTP) and established static web pages as content on the Web. Web 2.0 allowed users to collaborate and utilize server-side scripting to allow online services to proliferate. It is due to the growth of online services that led to the conceptualization of Web 3.0 being decentralized. Web 3.0 has been around as a concept since the early 2010s. The concept of Web 3.0 centers around user autonomy and not being reliant on centralized services, essentially having users be responsible for their data. The three generations of the Web can be seen in Fig.~ \ref{fig:webGeneration}.
     
     When looking at the contemporary state of Web 2.0 with its centralized system, it suffers from a large number of issues due to the impact of centralization. These issues include: 
     
     \textbf{Scalability and Availability}. Internet resources and services (e.g., computing, storage, network, and database resources ranged from single servers to large scale Cloud-based server-farms/datacenters) have limited capacity and cannot cope with the requirements of the increasing number of users without their direct contributions in providing resources. Large Internet companies may fail to provide resources to users in different geographical regions or over a specific time. This raises another issue, which is the availability of resources. In 2019, Microsoft Azure was reportedly running out of VMs for its customers in East U.S. \citep{zdnet}. Similarly, in March 2020, Azure has suffered from a shortage in data center capacity due to a large amount of demands resulted by Covid-19 pandemic \citep{covid}.
     
     \textbf{Reliability.} Services based on the client-server model are vulnerable to a single point of failure and bottleneck. They may fail to provide services due to problems like network or system failure. 
     
     \textbf{Security and Privacy.} Collecting user data by different service providers and storing them in a certain number of specific servers to support the hosting of various types of services and applications expose vulnerabilities and user data privacy to cybercriminals. 
     
     \textbf{Trustability.} Large Internet corporations and service providers are trusted parties that can potentially maintain, control, and administrate user data, access, and activities. While this can bring benefits for the users, it potentially can be used as a source of control to apply surveillance or censorship, or it can lead to abusing trustability~\citep{dinternet}. In this paper, we provide a systematic review of the potentials and capabilities of Blockchain-based solutions which can efficiently be used for any aspect of the Internet decentralization.  There are several other approaches for decentralizing the Internet, such as projects seen in Section \ref{types-of-decentralisation}. However, our focus in this paper centers around using Blockchain to decentralize the Internet. Also, it must be taken into account that Blockchain for IoT security is out of the scope of this paper due to space limitations.
     
     The rest of this paper organized as follows: Section \ref{sectionInternetArchitect} provides an overview of the current Internet architecture and what Blockchain is facing against, Section \ref{sectionUnderstandBC} revolves around understanding Blockchain's components and challenges it would face on decentralization, Section \ref{sectionConsensus} presents a list of consensus algorithms that have the potential to be a candidate for reaching consensus within the Internet, Section \ref{sectionFutureTech} discusses the emergence of future and old technologies that can be integrated with Blockchain, Section \ref{sectionDiscussion} presents a discussion of current and future technologies that can impact and benefit Blockchain in decentralizing the Internet, and finally Section \ref{sectionConclusion} concludes the paper and discusses future works.  

\section{Understanding the Contemporary Internet Architecture} \label{sectionInternetArchitect}

The Internet architecture has amassed to a tremendous scale where it encompasses many systems, services, protocols, architecture, and hardware to use on such an extensive scale. It is nearly impossible to cover every intricacy of the Internet. However, instead, this section mainly considers the macro-scale of the Internet's model and generally discusses why it is centralized and challenges caused by current centralized Internet architecture This section is  followed by summarizing the types of decentralization that can be achieved with the Internet. Only to finish on why Blockchain is the choice for decentralizing the Internet. 
 
     \subsection{Current Internet is Centralized} \label{subsectionCurrentInternetCentralized}
     
    The current Internet is centralized due to the unusual architecture that has been designed to route users to pass through a singular point before the users can interact on the Internet. This singular point on the Web can be seen in many forms, such as Domain Name System (DNS), where it acts as a translator for IP addresses and Domain names for human and computer readability. The DNS was implemented in a distributed way in Web 2.0, but traces of centralization are observed in domain name servers, namespace governance and operation \citep{blockDNS,socialDNS} . This centralization is further supported by the monopolization of DNS generation and distribution on the web by the Internet Corporation for Assigned Names and Numbers (ICANN) \citep{consortiumDNS}. Internet Service Providers (ISP) are another centralized point, as users need to establish a connection with an ISP before the user can interact with the Internet. This allows ISPs to have control over the Internet traffic and allow third-party organizations  to have access and control Internet traffic flow. The fact that the Internet is heavily dependent on DNS and ISPs to operate, proves our reasoning of centralization that occur within the Internet architecture.
     
     \subsection{Challenges of Centralized Internet} \label{sectionChallengesCentral}
    
    The Transmission Control Protocol/Internet Protocol \\(TCP/IP) is synonymous with the Internet when discussing how the Internet communicates ~\citep{internetArchitect}; this brings up the question of decentralization compatibility. Considering the fact that TCP/IP has been the catalyst for the Internet since the very beginning, most improvements to the Internet appear to be revolved around TCP/IP. 
    
    The Application layer is the standardizing layer to enforce applications, e.g., web-browsers and web-servers, running on end devices to conform to a regulated method to communicate with each other. HTTP and its secure successor Hypertext Transfer Protocol Secure(HTTPS) are well-known examples of protocols placed on the application layer. These protocols are the fundamentals of the Web, widely implemented over the Internet, and relying on the client-server networking model (a centralized architecture). Blockchain can be considered as an option to decentralize HTTPS. However, Blockchain is an entirely different system that communicates using its own standards and protocols, meaning that a communication method between HTTPS and Blockchain needs to be established. Furthermore, Blockchain employs completely different security measures compared to HTTPS. This difference in security lies in HTTPS using a multi handshake protocol ~\citep{deconstructTLS1.3}, while Blockchain uses a cross-referencing method.
    
    The transport layer encompasses communication protocols to provide end-to-end communication services such as reliability, traffic, and flow control to applications running on hosts ~\citep{internetArchitect}. Its services has been the same since early days of the Internet, which is to offer a connection between hosts. 
    
    Over the years, flaws and limitations have been uncovered for the transport layer, from enumeration attacks for extracting information about the targeted system and network, using fingerprinting techniques to uncover open ports of a system for infiltration, to SYN flood attacks to overwhelm a system. Due to this uncertainty of security created from the flaws in the transport layer, an alternative security solution should be sought. There are two options to partake in to resolve the problems presented in the transport layer. The first option revolves around \citep{thinSecurityLayerTCP/IP} by greenfieling and implement a policy-based security module into TCP/IP. This greenfield option would use four-way handshaking and public-key cryptography to ensure a secure entity that would maintain and monitor the security in the system. The second option would be to follow \citep{suitabilityOfBC} and brownfield it by transitioning TCP/IP into Named-Data-Network (NDN). NDN is the contending winner, as it is robust enough to offer enhanced performance for the network traffic. More of NDN will be discussed on the Section \ref{sectionFutureTech}. 
    
    The network layer is one of the major backbones with many inner mechanisms working in conjunction with this layer within the Internet architecture. This layer allows communication protocols such as IP for the delivery of packets since IP as by itself does not guarantee the delivery of the packets to the intended destinations. The network layer cooperates with the transport layer to deliver the packets via TCP, guaranteeing the arrival of packets on the destination node. Hosts on the Internet use names (i.e., domain names for the servers) or numbers (IP addresses for both servers and clients) or both of them to communicate across the Internet. Client hosts need to resolve server names to IP addresses before being able to initiate requests for communications. The DNS is an application layer service which is used to resolve names to addresses or vice versa. It is a vital protocol within the Internet model that translates unique IP addresses to human-readable addresses ~\citep{Design&ImplementationsOfDNS}. Security is an essential aspect in DNS, and methods of providing security such as extensions like Domain Name System Security Extensions (DNSSec) are used \citep{SurveyDNSSec} to mitigate against DDoS, configuration tampering, DNS poisoning, and information leakage as well as countless other DNS vulnerabilities\citep{SurveyDNSSec}. The DNS in the current Web 2.0 is centralized as discussed in Section \ref{subsectionCurrentInternetCentralized}; this brings up the question of how we decentralize DNS while maintaining the same functionality of translation. The solution to that centralization is a decentralized name system. The decentralization of the naming system can be seen with numerous proposals, where each employs Blockchain and Peer-to-Peer (P2P) technology to achieve decentralization. 
    
    SocialDNS \citep{socialDNS} employs short-names for resources in a localized area network while using a rank-based mechanism to handle name conflicts. SocialDNS uses P2P to enable virtual organization of the domain names, without the need of a central authority. BlockDNS \citep{blockDNS} is another solution for decentralizing the name system, as it allows users to apply domain names while maintaining authoritative server information in a decentralized way. BlockDNS employs the use of a lightweight verification system that can cut the overhead of data authenticity verification to a few hundred bytes, allowing the BlockDNS to handle more DNS queries in the DNS cache. ConsortiumDNS \citep{consortiumDNS} is another DNS to consider. It resolves the limitation of storage in a blockchain by using a three-layer architecture with external storage. This design in ConsortiumDNS allows indexing of transactions and Blockchain blocks for increased performance of domain name resolution. Last of the proposal is Bitforest, which uses a partially trusted centralized name server in a Blockchain with cryptocurrency to achieve decentralized trust and security.  Bitforest \citep{bitforest} is capable of the same performance and scalability of centralized Public Key Infrastructures (PKI) in client validation and verification of name bindings. Bitforest's architecture maintains decentralization by not allowing the administrator to violate identity retention.
     
    Internet Protocol Version 4 (IPv4) has an issue of address space limitation where it is not able to accommodate future IP addresses due to exhaustion of usable addresses \citep{ipv,comparativeIpv6Tunnel,IPv6Tunneling}. Internet Protocol Version 6 (IPv6) has always been seen to succeed IPv4, as it can solve the issue of address space limitation from the increase of the size of 32 bits to 128 bits \citep{ipv} and also solving many of the limitations and security issues within IPv4. Having a full transition from existing IPv4 to IPv6 is impossible due to the high cost of replacing existing IPv4 Internet infrastructures e.g., IPv4 routers. One important aspect within the arsenal of IPv6 is the ability of "tunneling" between IPv4 and IPv6. Tunneling allows IPv6 to encapsulate itself into an IPv4 address and cross-communicate with the existing IPv4 addresses \citep{fromTCP/IPtoConvergent}. The tunneling feature would be an essential component in the implementation of Blockchain for the Internet, as the Blockchain's domain consists of multiple interoperable smart contracts. Without this tunneling feature, nodes would only be able to communicate with IP versions that is supported. Not being able to communicate with other versions of IP would result in blocking off the other half of the Internet to communicate with. As of June 27th 2020, Google has collected statistics across the Web and have shown that 67.08\% of the web is still in IPv4, while the remaining 32.92\% has migrated to IPv6 \citep{ipv6statistic}. 
    
    Lastly, the data-link layer in the TCP/IP model consists of OSI's data-link layer and a physical layer. The physical part of this layer establishes the hardware need to establish interchangeability and interconnection of the network link between hosts, routers, and switches. The software part uses protocols to encapsulate packets received from the network layer into frames with Media Access Control (MAC) address and prep it to be ready for transmission. The data-link layer also provides synchronization and validation for the frames, as it transfers receiving packets with the corresponding and correct MAC address to the network layer. According to \citep{services&protocolDataLinkLayer}, the service this layer uses consists of WLAN, LAN, Ethernet, and other similar network devices to overcome the limitations of the network layer.
     
     The TCP/IP model may also include the 5th physical layer which  encompasses the hardware needed for sustaining the network \citep{internetArchitect,TCP/IP,securityTCP/IP}. This physical layer can be seen as a segregation of the data-link layer to establish clarity between hardware and software. However, this hardware layer could be prominent in the future with IoT, as computers are increasingly prevalent over time. 
     
     \begin{figure*}
         \centering
         \includegraphics[width=140mm]{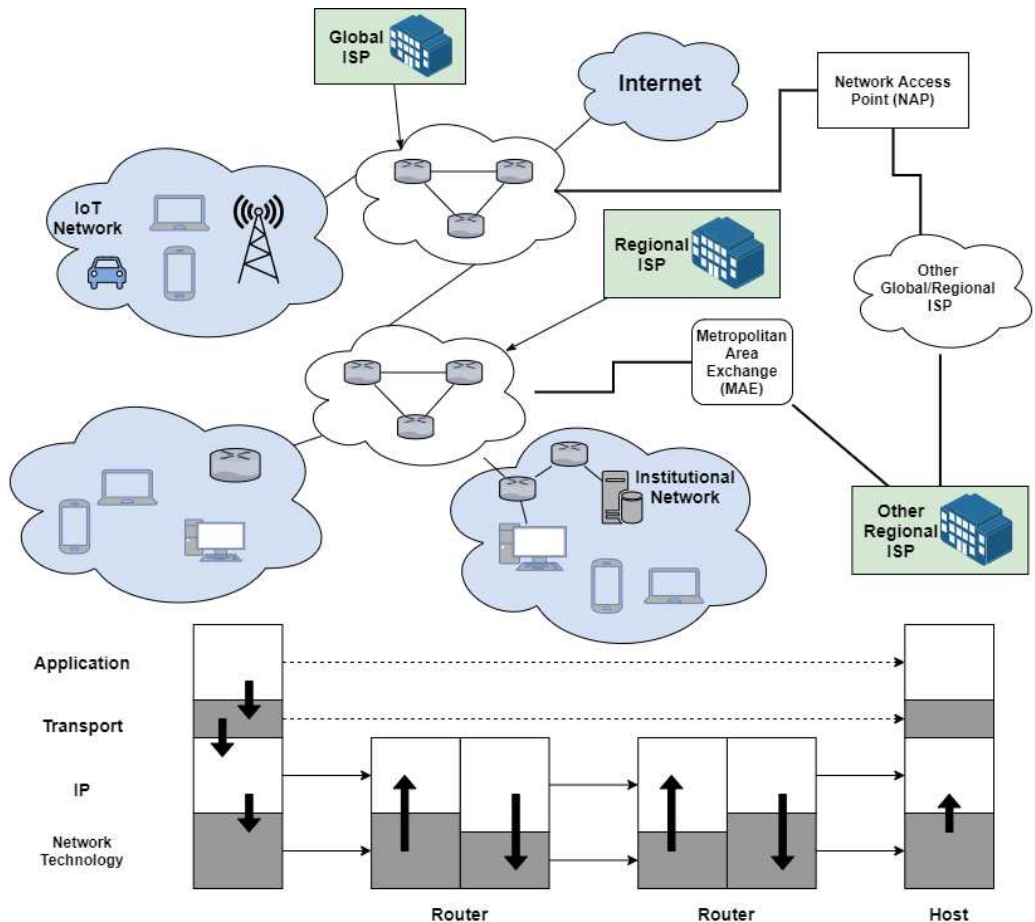}
         \caption{The Internet Architecture} \citep{bidgoli2004internet}
         \label{fig:internetArchitecture}
     \end{figure*}
     
     As illustrated in Fig.~ \ref{fig:internetArchitecture}, a centralized point is seen with each respective ISP. Users are provided access to the Internet through centrally administrated entities, which are so-called ISP networks. We understand that ISP plays the man-in-the-middle for computers to access the Internet, which resulted in this centralized Internet traffic route. A centralized infrastructure is always ideal in a private network for allowing a governing entity to easily administrate and have an overview of the network and its connections. There is also the case of fault tolerance systems where it accounts for preventing disruption on the network from a single component failure that has experienced prolonged continuity of operation. This allows for the lowered costs for IT equipment, expenditures, and maintenance. \citep{centralAgain} states that this lowered cost would enable the architecture to have a decreased level of maintainability and accommodate more expansion.
     
     Single point of failure is a major flaw in a centralized network, as it is caused by the need to trust for a central entity \citep{doNotDecentralize}. This singular point of failure can also be reflected as a singular point of control, where the central system can have total control of the network and its participating nodes. Security risks are another flaw, due to the possibility of compromised entry points into the infrastructure. These entry points would ensure a major risk for both the network and the databases. The second major flaw is caused due to the exponential growth of information data on the Internet. This exponential growth would cascade into the need for expanded capacity for data storage \citep{blockchainSurvey} to respond against the increasing information data. This need for increased storage data ties heavily to Big Data with extensive information data needing to be stored, resulting in a scalability issue. The scalability issues mainly come from using legacy databases that lack the efficiency and performance to respond to the ever-increasing information data needing to be stored across various devices on the Internet. \citep{blockchainToIoT} proposed implementing IoT into the Blockchain to cope with scalability issues by designing a new consensus algorithm that increases the throughput to handle the large information data, or locates the databases in a private or consortium Blockchain where it can process the database at a much higher speed. The current security with Alt-Svc that was introduced for HTTP \citep{alt-svcAbuse} has many underlying vulnerabilities such as bypassing black-listed sites, distributed port-scanning and DDoS of non HTTP sites. This makes Alt-Svc highly abusable for malicious purposes, and would be a critical issue. 
     
     \subsection{Types of Decentralization}
     \label{types-of-decentralisation}
     The original network design of the Web with HTTP by Tim Berners-Lee was envisioned to be a decentralized infrastructure \citep{redecentralize}. However, throughout its lifespan, as stated by \citep{web3.0BC},  the Internet has developed into a centralized infrastructure. The decentralized network option has been gaining traction, as the idea emphasizes on developing new protocols and underlying technologies through Peer-to-peer (P2P) technology for a shared data layer within the architecture \citep{web3.0BC}. A decentralized Internet would be able to give resiliency for data security, which would offer incentives for users to cooperate and further expand \citep{blockchainSurvey,blockchainToIoT}. This would increase scalability to support complex transactions of information data. Examples of decentralized Internet can be seen on projects like The Onion Route (TOR), Zeronet, and The Invisible Internet Project (I2P) \citep{TorVSI2P,zeronet,I2PCensorship}. The goal of these projects is to allow users to surf the Internet anonymously anywhere on the Internet while reducing their footprints. 
     
     Our research thus far states that there are two types of decentralized networks that can be achieved for a decentralized Internet. 
     The first being a completely decentralized network by \citep{doNotDecentralize} where "trust" and controls are spread across anonymous users. This "trust" is to ensure control is from a user and not from a centralized point. With the rise of personalized computers for the current generation of users, each computer would be following its configurations, not having a dictated standardization across all connected computers in the network. However, utilizing a fully decentralized network comes at the risk of losing the conveniences provided by Internet services that have been developed since Web 2.0.
     
     The other type of decentralization is the distributed network. A distributed network ensures that every participating computer is inter-connected and is co-dependent with each other. The inter-connectivity between computers would allow centralized legacy systems to run within the network in a decentralized way. To completely transition the Internet into a decentralized Internet, the distributed method would be a more plausible way of achieving it.
    
    \subsection{Blockchain for Decentralization}
    
      \begin{figure*}
          \centering
          \includegraphics[width=140mm]{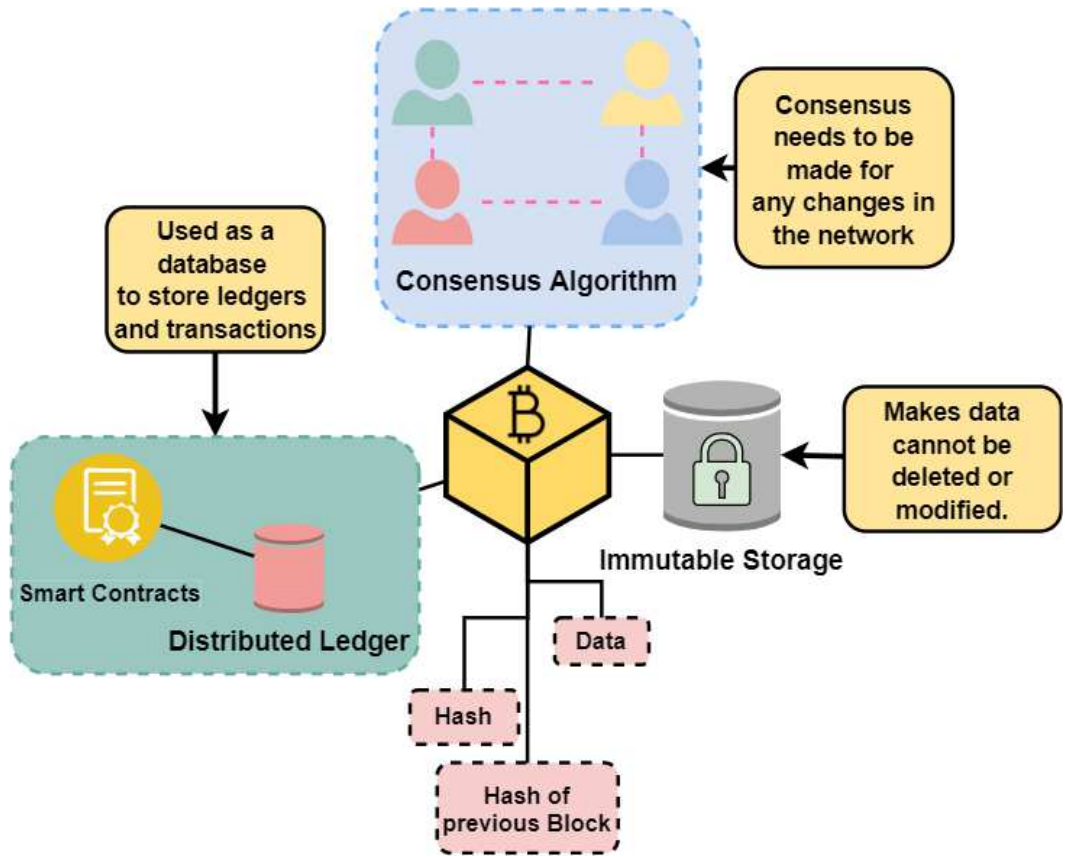}
          \caption{Blockchain Technology}
          \label{fig:blockchainTech}
      \end{figure*}
    
    Blockchain allows the Internet to achieve a distributed state of the network by allowing "trust" to be shared across the connecting networks. This "trust" gives the notion of web of trusts between nodes in the Blockchain. Furthermore, Blockchain has ties to the mentioned decentralized Internet projects in Section \ref{types-of-decentralisation}. Those projects have some peculiar traits, whereby P2P, data storage, and encryption play an essential role in each project. Blockchain also parallels these traits; therefore, we consider it as the most prominent option for Internet decentralization throughout this paper. The way Blockchain is able to accomplish these features is due to its components that are shown in Fig.~\ref{fig:blockchainTech}, which will be further explored in Section \ref{sectionUnderstandBC}.

\section{Understanding Blockchain-based Decentralization} \label{sectionUnderstandBC}
     \subsection{What is Blockchain?}
     
     \begin{figure*}
         \centering
         \includegraphics[width=140mm]{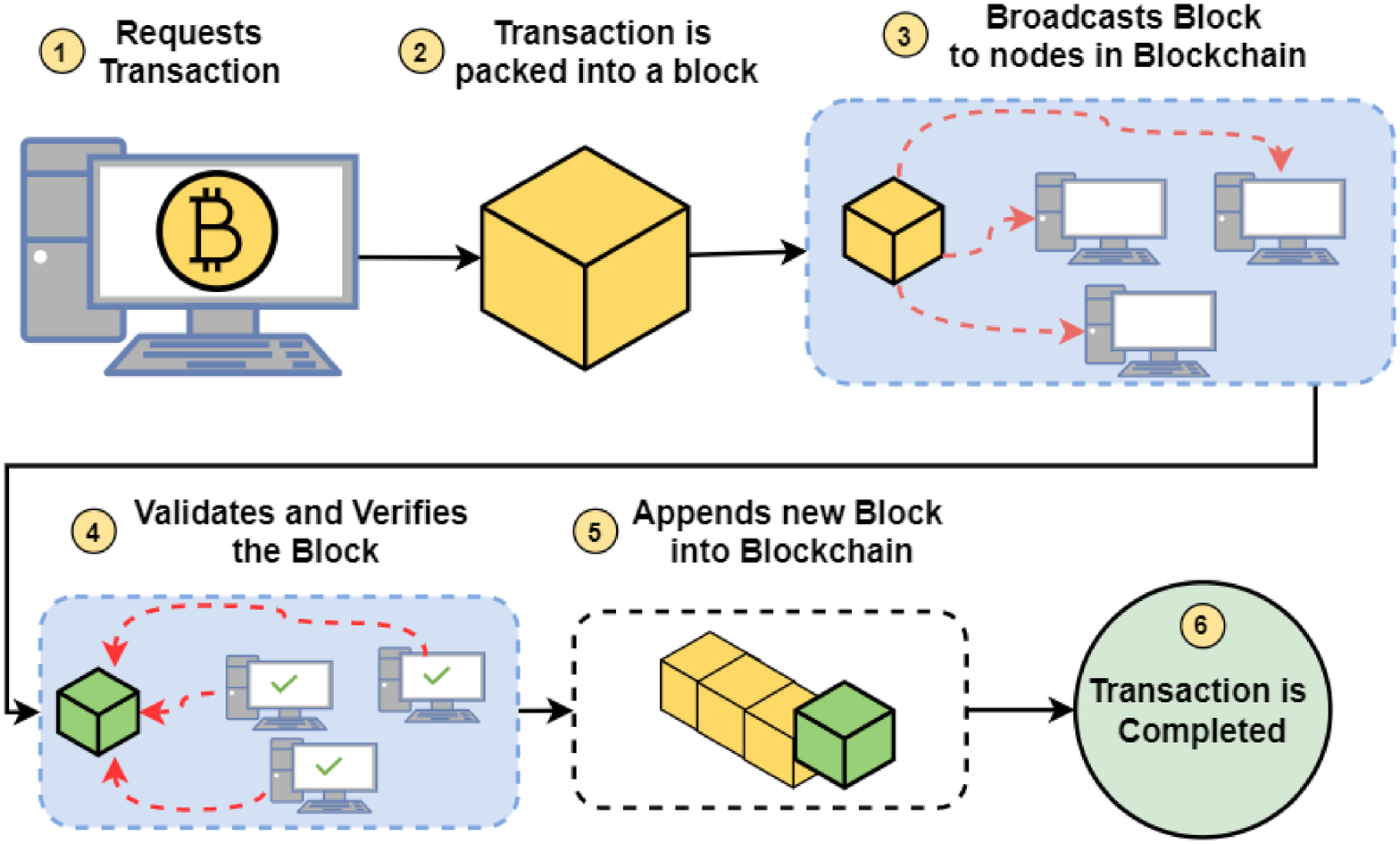}
         \caption{Blockchain Process}
         \label{fig:blockchainProcess}
     \end{figure*}
     
     Blockchain is described as a database that is used as a storage for a decentralized network \citep{whatIsBlockchain}. It is usually seen in its popularized usage on Bitcoin, Ethereum, Dogecoin, and other cryptocurrencies. The Blockchain is not limited within the boundaries of financial usage, as it can be expanded further upon to encompass other types of systems, applications, and make a decentralized network \citep{graphchainSemantic}. Asymmetric cryptography and distributed consensus algorithms are part of the systems within Blockchain, which provide user security and ledger consistency \citep{blockchainOverview}. In summary, Blockchain is a decentralized, and immutable database that facilitates its chain network with its participating nodes through a voting scheme. 
     
     As seen on Fig.~ \ref{fig:blockchainProcess}, where it demonstrates the overall Blockchain process, the process starts with the request of a transaction from a node, which would be packed into a block. This block is then broadcasted to other nodes within the Blockchain network for validation and verification. When that block has been successfully verified, it would be then appended onto the end of the Blockchain to be stored and finishing the transaction.
     
    Blockchain exhibits the following key  characteristics \citep{blockchainOverview,graphchainSemantic}:
     
     \textbf{Decentralization}, where each transaction through its peers in the network is done only by two nodes at a time and does not need a third-party validation. Decentralization allows the Blockchain to be non-reliant on a central authority. This  enables nodes to essentially have equal voting rights within the network, which is then utilized with the consensus algorithm to dictate the Blockchain.
     
     \textbf{Persistency}, where the transactions must be validated by trusted miners. Persistency ties into the technology of immutability to ensure the ledgers stored within the nodes are absolute and not modifiable nor be deleted.
     
     \textbf{Anonymity}, where each miner uses a generated address as a unique ID. Although not all Blockchains are anonymous and some practice pseudo-anonymity such as Ethereum. However, the core principle is to ensure miners within the network can remain anonymous.
     
     \textbf{Auditability}, where each transaction within the Blockchain always has a reference point that is imprinted into the nodes. This reference point is used to provide a trackable transaction that has been verified and enacted within the Blockchain. The auditability can be seen as the characterization of verification and trackability of the transaction in the Blockchain network.  
     
     It must be taken into account that the focus of this paper is on decentralization and the other aforementioned characteristics of Blockchain including persistency, anonymity, and auditability are out of the scope of this paper.
     
     \subsection{Components of Blockchain}
     Three main components run within the Blockchain system. All three components are required to work together, as these components give the pillars of support in ensuring decentralization for the Blockchain \citep{blockchainRevolution}. 
     
     \subsubsection{Distributed Ledger} \label{subsubsectionDLT}
     Distributed ledgers offers a distributed database \citep{blockchainRevolution} that form a network connection between users are called nodes. Within these nodes are ledgers, which are ordered list of transactions with timestamps. These ledgers can only be appended within the database \citep{HCIblockchain,distributedLedgerCompareAnalysis,analysisDLT,distributedLedger}, ensuring a secure way to track transactions without the need of a central figure for verification \citep{redecentralize}. Initial process of transactions were done in a P2P manner, only to be facilitated by Smart Contracts during the 2nd generation of Blockchains \citep{distributedLedgerCompareAnalysis}. Smart Contracts are software that controls the transmission of the ledgers between nodes. An alternative technology is that \citep{dontNeedLedger} proposed to replace distributed ledger using the browser as a lightweight middleware, but it is still in the testing phase.
     
     \subsubsection{Immutable Storage}
     The Immutable Storage is a component that refers to the nodes having the ability to be unalterable. Each database is retained in every node and has a reference of itself in the Blockchain as an immutable history \citep{blockchainOverview,HCIblockchain,immutabilityOfDifferentBlockchain}. Immutable Storage is a crucial component that provides encryption function to maintain the integrity of ledgers within the nodes. The Immutable Storage guarantees no other medium altering the content of the transaction, it would establish an increase of incentives and trust within the Blockchain.
     
     \subsubsection{Consensus Algorithm}
     The consensus algorithm is an algorithm for nodes to achieve consensus between nodes for alteration or modification of the existing ledgers \citep{HCIblockchain,demystifyBlockchain}, only to append them into a new block at the end of the chain within the Blockchain. The consensus algorithm moderates the Blockchain by dictating the nodes on how to achieve an agreement and update the Blockchain network \citep{HCIblockchain}. We further discuss Consensus Algorithm  in Section \ref{sectionConsensus}.

    \subsection{Types of Blockchains}
    
    Table \ref{table:propertiesBC} compares properties of three types of Blockchain, including public Blockchain, consortium (hybrid) Blockchain, and private Blockchain, for different criteria \citep{blockchainOverview,analysisOnConsensusProtocols,blockchainChallenge&Security,doYouNeedABlockchain}:

     \begin{table*}
     \centering
     \caption{Properties of Blockchain}
     \label{table:propertiesBC}
     \setlength{\tabcolsep}{3pt}
     \renewcommand{\arraystretch}{2}
     \begin{tabularx}{\textwidth}{|>{\centering\arraybackslash}X
     |>{\centering\arraybackslash}X
     |>{\centering\arraybackslash}X
     |>{\centering\arraybackslash}X
     |>{\centering\arraybackslash}X|}
     \hline
     \textbf{Properties}& 
     \textbf{Public Blockchain}& 
     \textbf{Consortium Blockchain}&
     \textbf{Private Blockchain}\\
     \hline
     Determination of Consensus & All miners & Selected nodes & An organization\\
     \hline
     Read Permission & Public & Public or Private & Public or Private\\
     \hline
     Immutability & Close to full immutability & Can be tampered & Can be tampered\\
     \hline
     Efficiency & Low & High & High\\
     \hline
     Decentralization & Yes & Partial Centralization & No\\
     \hline
     Consensus Process & Permissionless & Permissioned & Permissioned\\
     \hline
     Examples & Bitcoin (BTC), Ethereum (ETH) & Bankchain, R3 & Hyperledgers, PBFT, Quorum\\
     \hline
     \end{tabularx}
     \end{table*}
     
    \begin{itemize}
        \item Public Blockchain is opened for everybody to participate in the verification and consensus process within the Blockchain. The Public Blockchain is a permissionless Blockchain, where public users do not need special permission to join the Blockchain. Nodes in a Public Blockchain have full read and write permissions. Examples of Public Blockchain can be seen with Bitcoin and Ethereum. These cryptocurrency's development are open source, which can be viewed or modified by anybody.
         
        \item Consortium Blockchain only chooses selected nodes from a public or private branch of the Blockchain to handle the verification and consensus process in the Blockchain. As it is a hybrid between public and private Blockchain, it is labeled as a permissioned Blockchain due to applying the same logic of authorizing a few select nodes to have the read and write permission in the Blockchain. Examples of Consortium Blockchains are seen in the financial and health industry with Hashed Health and IBM/Maersk.
         
        \item Private Blockchain utilizes private nodes from an organization or group that is restricted from the public to handle the verification and consensus process of the Blockchain. Additionally, not every node can participate in both the processes, even if the nodes are from the same organization or group. The Private Blockchain is a permissioned Blockchain with the same principle of selected authoritative nodes, as it functions similarly to a Private Blockchain. The difference lies in that Consortium's authoritative nodes are not consolidated from a single group, but consist of multiple different groups. Examples of Private Blockchain are seen with Corda and Hyperledgers, where few nodes are only allowed modified.
    \end{itemize}
    
    Blockchains can be categorized into two groups in terms of user access. The permissionless Blockchain allows for open participation where every user has an equal vote (P2P). The permissioned Blockchain uses distributed mechanisms with a trusted third-party to have a shared mediating state between the exchanges of stakes. This permissioned Blockchain governs the consensus by restricting the access of the consensus protocol to the selected few governing nodes which can result in a centralized scenario.  However, there is a pressing issue with permissioned Blockchains formed from the dependency of the governing nodes that form the consensus. This issue creates an issue with trustworthiness, as nodes would need to trust these governing nodes to make the consensus for the Blockchain. However a permissionless Blockchain, in our opinion, would result in a lawless Blockchain where the consensus can be monopolized through majority votes. Despite the issue of dependability and trustworthiness in permissioned Blockchain, this can be solved by providing the governing nodes to be chosen in a decentralized and autonomous way~\citep{doNotDecentralize}
    
    \subsection{Generations of Blockchain} \label{subsectionGenBC}
    
        \begin{table*}
        \centering
        \caption{Generations of Blockchain}
        \label{table:blockchainGeneration}
        \setlength{\tabcolsep}{3pt}
        \renewcommand{\arraystretch}{2}
        \begin{tabularx}{\textwidth}{|>{\centering\arraybackslash}X
        |>{\centering\arraybackslash}X
        |>{\centering\arraybackslash}X
        |>{\centering\arraybackslash}X|}
        \hline
        \textbf{Properties} & \textbf{Gen 1} & \textbf{Gen 2} & \textbf{Gen 3}\\
        \hline
        Innovations & Distributed Ledger and Cryptocurrency & Smart Contracts, Decentralized Applications (dApp) and Digital Assets & Application in the industry\\
        \hline
        Design & Setup a shared public ledger to support cryptocurrency networks \& P2P & Security for transactions & Decentralized software architecture\\
        \hline
        Examples & Bitcoin, Litecoin, Dogecoin & Ethereum, Neo & IOTA, Holochain, Quarkchain\\
        \hline
        \end{tabularx}
        \end{table*}
        
    Blockchain is a developing technology, and developments in the next generation of Blockchain are already underway \citep{HCIblockchain}. The first-generation of Blockchain brought the concept of public ledgers for supporting a cryptocurreny network eco-system with PoW consensus. This concept gave us the creation of the first cryptocurreny with Blockchain, Bitcoin. The second-generation of Blockchain is rooted in cryptocurrency \citep{HCIblockchain} and brought the innovation of Smart Contracts, which was discussed in Section \ref{subsubsectionDLT}. There are already proposals of third-generation Blockchain in the market where it prioritizes providing support for different Blockchain data structures, interchain and intrachain proof protocols \citep{blockchainSurvey}. The applications of the third-generation Blockchain have evolved to a state where it can be considered as a decentralized software architecture, as it would have the scalability to handle large amounts of transactions with higher efficiency than the previous generations. The main attraction to achieving the decentralized Internet stems from the third-generation of Blockchain. The fourth-generation has not been clearly defined yet, as developments are prioritized in the contemporary third-generation of Blockchain. However what is being discussed in the community with the fourth-generation of Blockchain is that, the possibility of implementation with another technology such as AI or properties such as time. A proposal that is currently being developed as the fourth-generation of Blockchain is seen in SOOM, a developing Blockchain that utilizes time/space for increased security and processing speed.
    
    \subsection{Limitations of Blockchain} \label{subsectionLimitBC}
    
     Blockchain is not a fully decentralized system by design. It is considered as a partially decentralized system \citep{whatIsBlockchain}. There are simulations done on Blockchain where results have shown natural pressures of forming centralized nodes within the network \citep{blockchainOverview,graphchainBoyen,mastodon}. This slight centralization leads to a bigger picture of limitations and flaws inherent with the current second-generation of Blockchains. While Blockchain is a prominent emerging technology which has proved its efficiency in several areas, it also comes with its own set of challenges. These limitations and challenges include:

    \textbf{Scalability:}
        Each transaction is needed to be verified by a trusted central node~\citep{fromTCP/IPtoConvergent}, where the bottleneck would occur from the increasing transactions that are occurring every day \citep{blockchainOverview}. This is especially prominent in multichain Blockchains \citep{blockchainSurvey}. Multichain Blockchains are private Blockchains that are used for financial applications, where it would require the use of full hashes of the transactions. Multichain Blockchain's design is to ensure total security and control for the transactions, hence the need for using fully hashed transactions for communication. Using this full hashed transaction results in need for increased storage for communication in the network stream, where bottlenecks would heavily affect it. 
        
        All of the bottleneck issues stem from the scalability issue with blocksizes being limited to 7 transactions per second. However, this scalability issue is repairable through implementing relevant technologies like graphchain where parallel mining can be done to overcome the bottleneck \citep{improveBlockchainWithGraphchainAndParallelMining} and the implementation of edge computing and fog computing to further reduce the issue. \citep{monoxide} proposed the Chu-ko-nu Mining, a system which can bypass the scalability issue of limited transaction. Chu-ko-nu Mining introduced "Asynchronous Consensus Zones" where it uses multiple parallels and independent single-chain nodes to reduce communication and partition the workload of the transactions. Implementing this system would ensure mining across single-chain nodes be the same and deliver over a thousand times of throughput, and two thousand times of capacity compared to Bitcoin and Ethereum.

    \textbf{Performance:}
        The performance with the current generation of Blockchains is plagued with several issues that are making it slow and unscalable for large transactions. Smart Contracts has an issue of inefficient transmission between nodes, as \citep{blockchainSurvey} states that the Smart Contracts are not able to fully utilize arbitrary programs that are restricted by the immutability of specific blocks. The second issue is Forking, which is a split called "fork" formed from a Blockchain that has its block mined simultaneously by multiple nodes \citep{analysisOfForkOnEthereum}. \citep{onForks} states that Forking causes a network delay of more than 1000 seconds. Forking can also be exploited into a forking attack where back doors can be inserted into the new chain that was created from the divergent \citep{corkingByForking}. However, \citep{reducingForks} proposes its own PvScheme system where Forking can be mitigated, as it introduced a theory of probabilistic verification scheme to reduce the occurrence of forks. This theory with PvScheme is accomplished by not requiring verification of new blocks from each node in the Blockchain.  The third issue involves the performance bottleneck in Blockchain. This performance bottleneck is caused by long verification time from the blocksize's limited seven transactions \citep{blockchainOverview,blockchainSurvey}. To resolve this issue of a performance bottleneck, there would be a need to have an increased blocksize to house more storage. This blocksize increase can be expanded with the proposed "Layer 2" system protocol with Lightning Network. An alternate solution is to harness Forking to allow more transactions.
        
    \textbf{Privacy:}
        Although Blockchain's innate security provides anonymity for the user by hashing the public key and private key, there have been findings by \citep{blockchainChallenge&Security} where both keys can be compromised. Both embedded keys can be extracted to show user's private information \citep{blockchainOverview}. The keys can be further exploited into erasing stored information data in the nodes \citep{eraseDataFromNode}. It is also possible to trace the user's address to the identities of users that execute transactions.  This identity tracing is caused by the nodes using the same false address continuously, as the Blockchain does not refresh a new false address for the node \citep{blockchainOverview}.
        
    \textbf{Mining Issue:}
        Selfish mining is a major issue within the Blockchain, as selfish miners would store their mined blocks. These mined blocks are released only after the selfish miner's requirements are met. Selfish mining would cause wastage of resources by the normal miners for mining blocks, as selfish miners would have a private branch that may have shorter chains than the public branch of the chain \citep{blockchainOverview}. Personalization mining is another issue in the Blockchain which comes from being unable to specify Blockchains to interact with Internet services. These mining issues can be solved by making parts of the Blockchain smarter with artificial intelligence to reduce the likelihood of personalized mining \citep{blockchainSurvey}.

\section{Investigating the Consensus Algorithm in Blockchain for Decentralization} \label{sectionConsensus}

     \citep{impossibleDecentral} states that a good decentralized Blockchain depends on a good consensus algorithm. A reliable decentralized consensus algorithm should not rely on trusted third-party services \citep{impossibleDecentral}, leading to the dismissal of permissionless Blockchain as a choice. The permissioned Blockchains is the favoring choice, due to being able to give both dependability and trust in a decentralized way \citep{doNotDecentralize}. There is also the matter of fog computing and edge architecture to account for, as it has relevance to IoT and the Internet infrastructure in terms of providing performance without latency issues for nodes connected at the "edge" of the Blockchain network. All of these variables give us the reasoning for needing to explore the available consensus algorithms.

     There is a variety of consensus algorithms in the current market to select from, with new ones being developed. Suggestions can only be made for consensus algorithms due to the uncertainty of these algorithms. The following sections would cover the selected consensus protocol and review how compatible it would be for the Internet. A table consisting of the consensus algorithms that have been reviewed is done in Table \ref{table:proofConsensus} and Table \ref{table:bftCrashConsensus}.

\subsection{Proof based Consensus Algorithm} 

     Proof based consensus revolves around nodes competing with each other to calculate and solve a cryptographic problem. Whoever solves the problem will earn the right to append the Blockchain. After appending the Blockchain, the cycle restarts. This type of consensus is widely seen in permissionless Blockchains \citep{consensusAlgorithmSurvey}.
        
        \begin{table*}
        \centering
        \caption{Comparison of Proof-based Consensus Algorithm}
        \label{table:proofConsensus}
        \setlength{\tabcolsep}{3pt}
        \begin{tabularx}{\textwidth}{|>{\centering\arraybackslash}X
        |>{\centering\arraybackslash}X
        |>{\centering\arraybackslash}X
        |>{\centering\arraybackslash}X
        |>{\centering\arraybackslash}X
        |>{\centering\arraybackslash}X
        |>{\centering\arraybackslash}X|}
        \hline
        \textbf{Consensus Algorithm} &
        \textbf{Blockchain Type} &
        \textbf{Permission Type} &
        \textbf{Decentralization} &
        \textbf{IoT Suitability} &
        \textbf{Efficiency for DI} &
        \textbf{Remarks}\\
        \hline
        PoW (Work) & Public \& Private & Permissioned & Medium & Yes & High & High Computing Power
        Wastage\\
        \hline
        PoET & Consortium \& Private & Permissioned \& Permissionless & Medium & Yes & Medium & Dependent on Intel's SGX\\
        \hline
        PoS (Search) & Private & Permissioned & Low & Plausible & High & Dependent on resource provision\\
        \hline
        PoAh & Public & Permissioned \& Permissionless & High & Yes & High & Low computation need when implemented with fog and edge computing\\
        \hline
        PoP & Public & Permissionless & High & Plausible & High & Requires further research\\
        \hline
        PoS (Stake), LPoS, dPos & Private & Permissioned & Low & Medium to High & Plausible & Requires further research\\
        \hline
        PoI & Public & Permissionless & High & Plausible & Medium & Requires further improvements\\
        \hline
        PoB & Public & Permissionless & High & No & Low & Requires monetary value\\
        \hline
        PoC & Private & Permissioned & Medium & No & Medium & Uses Storage as mining rights\\
        \hline
        PoA (Activity) & Public & Permissionless & High & No & Low & Can experience high levels of Delay\\
        \hline
        PoW (Weight) & Consortium \& Public & Permissioned \& Permissionless  & Medium & Plausible & Low & Requires monetary values\\
        \hline
        Casper & Consortium \& Public & Permissioned \& Permissionless  & High & No & Medium & Unable to meet IoT requirements\\
        \hline
        PoL & Public & Permissionless & High & No & Medium & Efficiency not high enough for IoT\\
        \hline
        \multicolumn{3}{p{200pt}}
        {\textbf{IoT Suitability}:Level of compatibility with IoT,
        \textbf{Efficiency for DI}: Level of efficiency in achieving decentralization}
        \end{tabularx}
        \end{table*}

            \subsubsection {Proof-Of-Work (PoW)}
            Widely used in a lot of Blockchain \citep{demystifyBlockchain}, PoW has its foundation from cryptocurrencies like Bitcoin and Ethereum. PoW uses computational power competition between nodes in solving a mathematical puzzle \citep{consensusAlgorithmSurvey}. For each round of consensus, the winner is given both rewards and power to create the next block in the Blockchain \citep{consensusAlgorithmReview,analysisOnConsensusProtocols,consensusAlgorithmComparativeAnalysis}. A new round would start, increasing the Blockchain indefinitely. PoW has a major flaw where it causes huge wastage of power for the calculation \citep{blockchainOverview}. This wastage of power extends to IoT devices being unable to compete with high computing power \citep{overviewBlockchainIoT}. The complexity of the calculation is determined by the overall computational power of the Blockchain \citep{consensusAlgorithmComparativeAnalysis}, and the length of the chain is proportional to the amount of workload \citep{consensusAlgorithmReview}. All of these flaws of power wastage and high computational power requirement makes PoW not optimized enough to be chosen for reaching consensus in a Blockchain.
            
            \subsubsection {Proof-Of-Elapsed Time (PoET)}
            PoET is a consensus algorithm that functions similarly to PoW, where it requires computation power to solve a calculation to create the next block. PoET differs from PoW, where there is no competition between stakeholders in solving the calculation. A winner is chosen based on whoever expires first from a random waiting time. PoET also has a considerably lower need for power consumption and sports a low latency and high throughput, making it a potential protocol for the decentralized Internet and particularly for IoT devices with limited resources \citep{surveyOnConsensusIoT}. Although an issue arises, as PoET's verification process is dependent on Intel's Software Guard Extension (SGX) \citep{Consensus-PoL}, thus making the consensus protocol having a centralized point, hence defeating the purpose of being a decentralized network.
            
            \subsubsection {Proof-Of-Search (PoS)}
            PoS uses the wasted power formed from PoW to calculate and give optimization solution for the Blockchain \citep{proofOfSearch}. The PoS is designed to offer computational service within a grid computing infrastructure, which is suited for large networks like data centers. However, the PoS process requires each node to check large amounts of possible optimization solutions. This might present a problem with large computation requirements, where it would hinder the performance and compatibility with IoT. 
            
            \subsubsection {Proof-Of-Authentication (PoAh)}
            PoAh is a consensus algorithm that targets IoT \citep{proofOfAuth}. PoAh removes the reverse hashing function in favor of utilizing an energy-efficient lightweight block verification method. The verification process of PoAh would authenticate the block and the source of the block. A node gains a trust value after completed a verified transaction. The trust value is a core part of the PoAh consensus protocol. PoAh is also scalable enough to integrate fog computing and edge infrastructure, due to its efficient verification. For PoAH to be able to benefit from future technologies while maintaining a lightweight consensus method, makes PoAh to be a viable consensus protocol.
            
            \subsubsection {Proof-Of-Property (PoP)} 
            PoP is a lightweight and scalable consensus protocol that provides "proof" for properties within the data structures of Blockchain \citep{proofOfProperty}. This "proof" is tied to the unique addresses of the node. The "proof" stores the state of the Blockchain in every newly created block, which is a concept from Ethereum's design. PoP is energy-efficient due to the "proof" design that allows the nodes to lessen the amount of information needed for every transaction. PoP would be a possible candidate for usage in IoT due to its reduced storage and processing power needed to join the Blockchain. However, PoP has not yet been successfully applied in the industry and requires more time to be developed. Thus, making PoP not a choice due to its infancy phase.
            
            \subsubsection {Other Proof-based Consensus}
            Despite many consensus algorithms to pick from, there is also a list of consensus algorithms that fall in the latter categories of not applicable. Such categories of consensus algorithm have gimmicks such as depending on specific data like cryptocurrency to function or depending on a node that has the most active hour in the Blockchain. This need for features within the consensus algorithm is seen as not desirable in the Internet architecture, as it only creates more complex transactions that will have no benefits. Consensus algorithms like Proof-Of-Stake (PoS), and its variants of Leased Proof-Of-Stake (LPoS) and Delegated Proof-Of-Stake (DPoS) are dependent on the usage of monetary values like cryptocurrencies as a stake. These three protocols require further development before it can be used practically in the Blockchain \citep{surveyOnConsensusIoT,reviewExistingConsensus}. There are also other consensus protocols that revolve around the need for utilizing monetary concept as well, Proof-Of-Importance (PoI) where it prioritizes nodes with more activity in the network which can potentially be adapted but needs more research \citep{surveyOnConsensusIoT,consensusAlgorithmReview}, Proof-Of-Burn (PoB) where it uses the concept of burning monetary values, Proof-Of-Capacity (PoC) where it requires a large volume of storage \citep{surveyOnConsensusIoT,reviewExistingConsensus}, Proof-Of-Activity (PoA) whereby it can experience a higher level of delay which is not suitable for delay-sensitive computers \citep{reviewExistingConsensus,proofOfActivity}, Proof-Of-Weight (PoW) where it depends on the amount of crypto coins a stakeholder possesses \citep{reviewExistingConsensus}, Casper which is an adaptation of PoS but is incapable of handling challenges that are present in IoT \citep{surveyOnConsensusIoT}, and lastly Proof-Of-Luck where despite its system being fully randomized and energy-efficient, its computation efficiency is not high enough to accommodate for IoT \citep{Consensus-PoL}.
    
        \subsection{Voting (Byzantine-based) Consensus}
            The concept of the Byzantine based consensus revolves around tackling the concept of the Byzantine General Problem, whereby in Blockchain's scenario, a node may fail and return leading false messages for the system and user \citep{reviewExistingConsensus}. This concept is usually referred to as the Byzantine Fault Tolerance (BFT) when used as an algorithm. The Byzantine-based Consensus takes into account of false leads or voting in the voting process when reaching consensus. 
            
            \begin{table*}[hbt!]
            \centering
            \caption{Comparison of BFT and Crash-based Consensus Algorithm}
            \label{table:bftCrashConsensus}
            \setlength{\tabcolsep}{3pt}
            \begin{tabularx}{\textwidth}{|>{\centering\arraybackslash}X
            |>{\centering\arraybackslash}X
            |>{\centering\arraybackslash}X
            |>{\centering\arraybackslash}X
            |>{\centering\arraybackslash}X
            |>{\centering\arraybackslash}X
            |>{\centering\arraybackslash}X|}
            \hline
            \textbf{Consensus Algorithm} &
            \textbf{Blockchain Type} &
            \textbf{Permission Type} &
            \textbf{Decentralization} &
            \textbf{IoT Suitability} &
            \textbf{Efficiency for DI} &
            \textbf{Remarks} \\
            \hline
            PBFT & Private & Permissioned & Medium & Yes & High & Limited scalability\\
            \hline
            dBFT & Private & Permissioned & Medium & Plausible & Low & Suffers with low network speed\\
            \hline
            SCP \& Ripple & Public & Permissionless & High & Plausible & Medium & Suffers with latency issues\\
            \hline
            Hyperledger \& Variants & Private & Permissioned & Low & Mostly No & Low to Medium & Requires further improvements for a lot of the variants\\
            \hline
            PoA (Authority) & Private & Permissionless & Low & No & Medium & Conflicting design methodology\\
            \hline
            Paxos & Private & Permissioned & Low & No & High & Needs to be adapted\\
            \hline
            Raft & Private & Permissioned & Medium & Plausible & High & Requires further improvements\\
            \hline
            \multicolumn{3}{p{200pt}}
            {\textbf{IoT Suitability}:Level of compatibility with IoT,
            \textbf{Efficiency for DI}: Level of efficiency in achieving decentralization}
            \end{tabularx}
            \end{table*}

                \subsubsection {Practical Byzantine Fault Tolerance (PBFT)}
                PBFT was the first system from 1999 proposed to solve a transmission error with its efficient algorithm \citep{consensusAlgorithmReview} where it provides high throughput, low latency, and lowered power usage as compared to PoW \citep{studyOfBCDecentralisedConsensusAlgorithm}. This results in PBFT being favorable for IoT networks \citep{surveyOnConsensusIoT}. PBFT requires all the nodes to take part in the consensus process, and only need 2/3rd of all node's agreement to reach consensus. However, it lacks scalability to work in a permissionless Blockchain due to its limited scalability caused by high network overhead and a low tolerance for exploits \citep{surveyOnConsensusIoT}.

                \subsubsection {Delegated Byzantine Fault Tolerance (dBFT)}
                The dBFT applies similarly to PBFT with the addition of not requiring the participation of all nodes, rendering it more scalable than its predecessor. A quirk with dBFT is that certain nodes are chosen to represent others or a group of nodes. Despite the scalability improvement, the network performance is not within an acceptable range due to its 15 seconds of average latency for creating new blocks in the Blockchain \citep{surveyOnConsensusIoT}. Thus, making dBFT not suitable as a candidate for reaching consensus in the Blockchain due to its slow performance.
                
                \subsubsection {Stellar Consensus Protocol (SCP)}
                SCP uses a variant of PBFT called Federated Byzantine Fault Tolerance (FBFT) and is a publicly opened decentralized protocol \citep{surveyOnConsensusIoT}. SCP allows complete "freedom" for the nodes to trust one another. This "freedom" of trust is used for assisting the process of reaching consensus. SCP calls a set of nodes a quorum, and a quorum is made up of multiple quorum slices. A quorum slice represents the trust between nodes. This binding of quorum slice forms a web-like structure in a P2P fashion \citep{consensusAlgorithmSurvey}. SCP can offer both high throughput and low power usage but suffers from latency issues caused by significant  network overhead. There is also a lack of security for the specific scenario of selecting an incorrect quorum slice to connect. Both of these issues cause SCP to be not suitable for reaching consensus.
                
                An alternate to SCP is Ripple, which is capable of reducing the latency for the Blockchain. Ripple can tolerate up to 20\% of faulty nodes. Despite the focus on solving the latency issue, Ripple is aimed for monetary purposes and is not fast enough for IoT \citep{surveyOnConsensusIoT}.
                
                \subsubsection {Hyperledgers}
                Hyperledger is a series of open-source Blockchain projects \citep{blockchainChallenge&Application} that has huge backing from big technology providers such as Linux and Intel. Certain projects within Hyperledger does have interesting options to consider. These Hyperledger projects are aimed directly at permissioned Blockchains. Hyperledger Fabric is a distributed ledger protocol that is run by peers within the Blockchain \citep{consensusAlgorithmSurvey}. However, the design of Hyperledger Fabric, even as of now in version 2.0, operates in a distributed manner with certain aspects needing certifications created by a centralized point with a Smart Contract called Chaincode \citep{comparativeAnalysisBCProblems&Recommendations}. This makes Hyperledger Fabric not ideal for decentralization due to its dependence on a singular service. Hyperledger Sawtooth is still in its infancy stage, as it requires more development before it can be taken into consideration. Hyperledger Indy has a lack of notable features to be used as a use case. Hyperledger Burrow has an issue where networks may halt due to the lack of specific roles within the Blockchain \citep{surveyOnConsensusIoT} as the Hyperledgers needs a "leader" within the permissioned Blockchain to reach a consensus \citep{consensusAlgorithmSurvey}, making it not suitable for reaching consensus in the Blockchain with its reliance of a leader. Hyperledger Iroha might instill some promises with its mobile design, making it compatible with IoT. 
            
                \subsubsection {Proof-Of-Authority (PoA)}
                Despite it being part of the Proof based consensus protocols, its design is based on BFT. PoA is a solution for solving PoW's issue of high latency, low transaction rate, and power wastage. PoA is designed to restrict the creation of new blocks to a fixed set of nodes that are selected with the Byzantine method \citep{proofOfAuthorityMDP}. This restriction of creating new blocks makes PoA designed for an enclosed network system with an administrator. The need for an enclosed network and an administrator in PoA, makes it not a suitable choice for reaching consensus in the Blockchain, considering everybody should have access to the Internet.
                
            \subsection{Voting (Crash-based) Consensus}
            The crash-based consensus algorithm is a sub-category of Byzantine-based consensus that tackles "crash failure". This "crash failure" refers to a crashed node not being able to recover by itself. But these crashed nodes are taken into account when reaching consensus. Unlike the Byzantine-based, this type of consensus is not capable of sustaining a full 100\% crash tolerance for the Blockchain system. 
            
                \subsubsection {Paxos}
               Paxos is a highly theoretical consensus algorithm that was one of the first few consensus protocols that were proposed \citep{studyOfBCDecentralisedConsensusAlgorithm}. Due to its theoretical nature, Paxos is challenging to understand and implement as a system \citep{reviewExistingConsensus}. Paxos has a crash tolerance level of up to 50\% \citep{surveyOnConsensusIoT}, hence why it is a crash-based consensus algorithm. Paxos was designed for smaller enclosed networks, which makes it not suitable for Internet implementation. However Paxos's safety feature in its balloting and anchoring system would be useful for the Internet and IoT \citep{paxos&BC}. Paxos's design comprises of two main roles, the leader and the follower. Depending on different documentations, there are as many as five roles in Paxos. The leader is chosen by the follower's ballot and makes progress within the protocol. The follower acknowledges the leader and provides its vote to the leader. A major issue lies in the leader role dominating the follower role. This issue makes Paxos run in a centralized-like way despite the possibility of being implemented in a distributed way.
                
                \subsubsection {Raft}
                The Raft algorithm is an attempt on trying to make Paxos more accessible and easier to understand and implement \citep{surveyOnConsensusIoT,studyOfBCDecentralisedConsensusAlgorithm}. Raft achieves the same effect and efficiency of Paxos, but with a lower crash tolerance level of 40\% \citep{consensusAlgorithmSurvey}. Since Raft follows a similar architecture of Paxos, this results in the same issue of a dominating centralized leader role.
                
\subsection{Usability of Consensus Algorithm}

    Three possible candidates which are PoP, Paxos, and PoAh stood out as potential candidates for the method of consensus that is suitable for an Internet architecture. PoP has a reduced need for storage and processing power. It is a strong contender due to its association with semantic technology for providing identities to properties of data structures. However, PoP suffers from a lack of practical testing, making it requires further development. Paxos is the second choice due to its potential for applicability. It has a history of being adapted to a wide array of systems, making it highly reputable for repurposing. Nevertheless, Paxos suffers from the difficulty of understanding its protocol and implementability. Making Paxos a plausible solution, but requiring a development team to modify Paxos for Blockchain. Finally, PoAh fits the criteria of being robust, scalable, and secure enough to handle the Internet, IoT, fog computing, and edge infrastructure. PoAh's trust system is an effective tool for establishing trustworthy nodes to interact on the Internet while maintaining equal voting power between all nodes.  All these factors make PoAh a suitable candidate for decentralizing the Internet. The identification of the consensus algorithm for the decentralization of the Internet architecture would provide the needed protocol for ensuring decentralization between roles. What is left to consider is Blockchain with its limitation from Section \ref{subsectionLimitBC} and how it can be resolved by incorporating other emerging technologies.

\section{Blockchain and Future Internet Technologies} \label{sectionFutureTech}
     
     \begin{figure*}
         \centering
         \includegraphics[width=140mm]{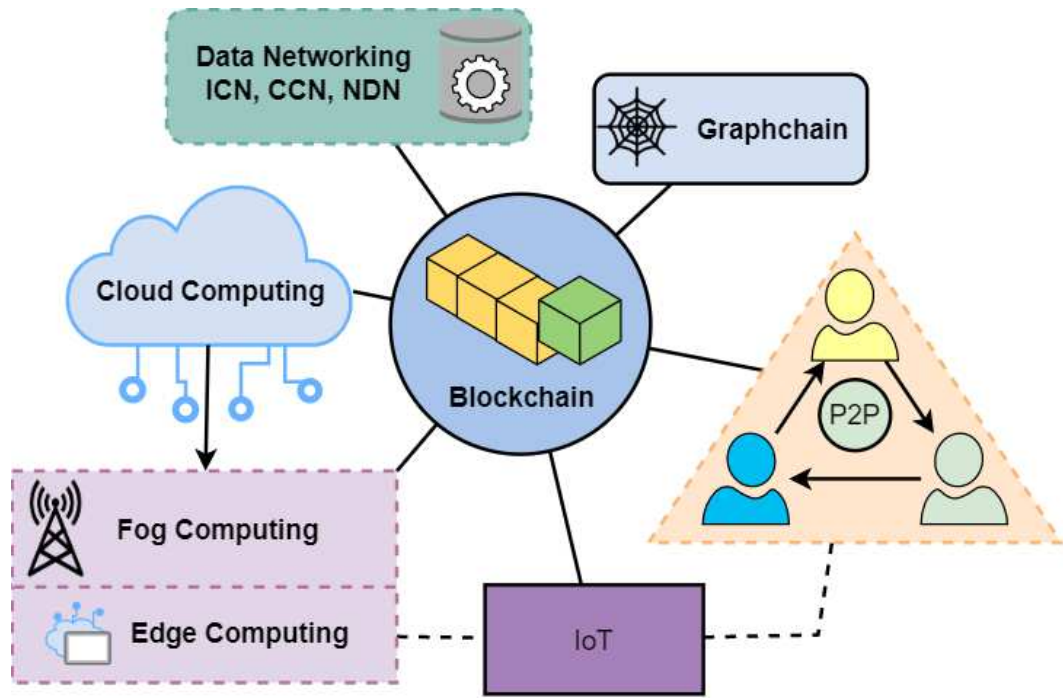}
         \caption{The Relation between Blockchain and Future Technologies}
         \label{fig:blockchainFuture}
     \end{figure*}

   New technologies are constantly evolving  and introduced every day, creating new integration opportunities for further improvement of Blockchain. The Internet has implemented a multitude of systems and protocols to work together and developed to become what it is today. Blockchain can adopt that same method of Internet's development by incorporating other technologies to work together and  improve Blockchain as a system. IoT has been an increasing presence within the industry, making IoT a relevant technology that would impact the Blockchain's participating node's hardware requirement. Since its conception, Cloud Computing has been an effective network and resource sharing technology, making it ideal for Blockchain with its resource allocation and bridging Blockchain to IoT. Graphchain is a developing technology that improves Blockchain, making it a possible alternate solution with a Graphchain-based Internet. Edge Computing and Fog Computing are technologies that enhance Cloud Computing by providing equal performance for nodes connected at the "edge" of the Internet. P2P technology is associated with the early days of file-sharing technology, making it vital to understand the sharing of resources between peers in a Blockchain. Lastly, Data Networking covers possible architectures that can replace the current TCP/IP architecture and change how information data would be connected. All of the topics that will be covered in this section is illustrated in the Fig.~\ref{fig:blockchainFuture}.      
   
\subsection{Internet Of Things (IoT)}

    IoT has established a new standard for today's technology by pushing the connectivity of the Internet to smart devices. \citep{surveyOnIoT} states that this new standard made smart devices centralized into a massive architecture. However, implementing IoT into Blockchain would expand how a node can take part in the Blockchain. This expansion is achievable with an increase of smart devices that can replace traditional desktop computers as a node. This expansion also provides an increase in scalability for Blockchain. IoT has a major challenge of needing multiple different devices to act as different main-in-the-middle for operations within the network \citep{challengesIoT}. Since there are no communication standards for IoT between different types of smart devices, this could lead to limitations of storage and computation power. This challenge would introduce the need for dedicated servers and infrastructure catered for IoT devices. But this challenge can be overcome with the implementation of cloud computing's resource provisioning.
    
    Research has been done for implementation of Blockchain \citep{BCintegrationWithIoT,IoTBClitReview}.  However, the research is flawed due to its test cases using cryptocurrency reliant blockchains and consensus algorithms. Since IoT will be a key technology that is already in the process of becoming the new norm, it would be crucial to implement IoT into Blockchain. Nevertheless, current implementation methodologies would need further research before it can be properly integrated.

\subsection{Cloud Computing}
    Cloud computing has become the new norm for today's generation of network technology with its power of resource pooling and virtualization. IoT has many similarities with Cloud computing since both principal centers around increasing efficiency for the network to operate. Cloud computing would also solve IoT's challenges \citep{ccAssistedBCEnabledIoT,blockcloud}. By implementing IoT and Cloud computing, it would increase the efficiency and performance of verification processing for the nodes in the Blockchain. The integration of Cloud computing by itself would also increase security, scalability, and the lowering of data storage for transactions in the Blockchain. Much of these efficiency increases could allow the integration of more types of consensus algorithms with lowered requirements of data storage and network overhead. \citep{surveyFog} states that current cloud computing is used in a distributed state that is composed of multiple components within the network, where it used to maintain fail-safe protocols. Blockchain technology can inversely help cloud computing's security by making the cloud storage's information and data be immutable, persistent, and decentralized \citep{BCinCCtoOvercomeSecurityVuln,securityImplicationOfBC&Cloud}. 
    
    A study done by \citep{integratedBC&Edge} have shown the inevitability of Blockchain, cloud computing, and IoT converging into Blockchain-of-Things (BCoT) as a form of evolution and become a future infrastructure. Cloud computing has the potential to be adapted as a service for the future Internet \citep{ccEnableFutureIoS}. However, there is an issue with the communication protocol between cloud computing, IoT, and Blockchain. There are no standardized communication protocols for all three technologies to communicate with each other \citep{onTheIntegrationOfCC&IoT}. This makes the development of the protocol a priority before it can be integrated with Blockchain or IoT.

\subsection{Graphchain}
    \citep{improveBlockchainWithGraphchainAndParallelMining} states that Graphchain is a technology that replaces Blockchain's network structure with a graph data structure. Graphchain is considered an improved version of Blockchain. Graphchain uses the same components of Blockchain but differs from Blockchain's linear chain by becoming a decentralized graph of self-scaling and self-regulated cross-verifying transaction framework \citep{graphchainBoyen}.  Graphchain disseminates the transaction data in "data shards" between multiple nodes in the graph chain, rendering it effectively scalable with high-performance. Graphchain also has the benefit of using parallel mining \citep{improveBlockchainWithGraphchainAndParallelMining} for increased performance and transaction processing. Graphchain is capable of being implemented with semantic technology of providing relations and "meaning" for the data structure to enhance the distributed ledger component \citep{graphchainSemantic}. But there is an issue with Graphchain where despite the necessary step to assure a decentralized system, centralization occurs within Graphchain due to a common descendant being shared between all newly created transactions \citep{graphchainBoyen}. This centralization issue pales in comparison to the benefits Graphchain would provide for the decentralized Internet.

\subsection{Edge Computing}
     
     \begin{figure}
         \centering
         \includegraphics[width=70mm]{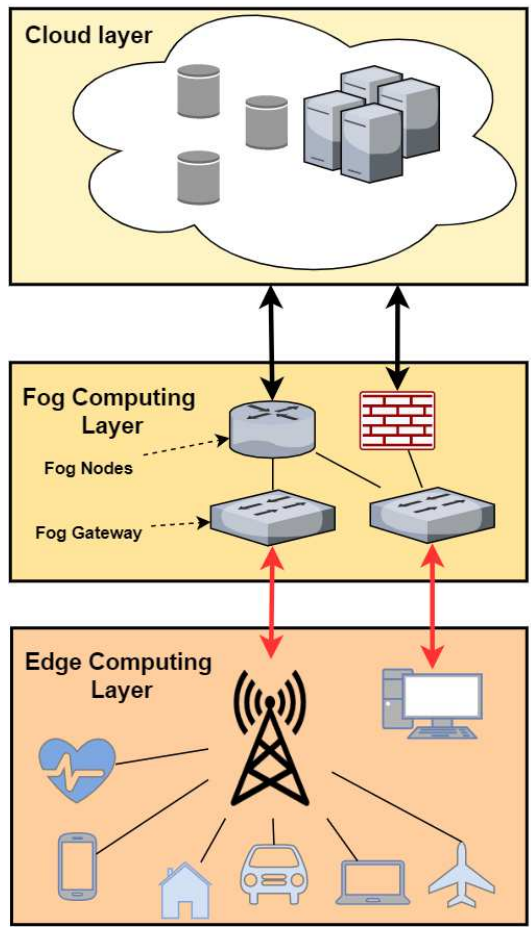}
         \caption{The Relation between Fog Computing and Edge Computing}
         \label{fig:FogAndEdge}
     \end{figure}
    
     Edge computing is a system designed by Cisco in 2014 to expand cloud computing by distributing cloud resources to the "edge" of the cloud network, forming an "edge" cloud \citep{integratedBC&Edge}. Edge computing centers around the concept of reaching the "edge" of the network. Edge computing operates similarly to Fog computing, as both technologies give benefits of scalability, security, and performance. Both edge computing and fog computing's interaction can be seen in Fig.~\ref{fig:FogAndEdge} \citep{futureEdgeForIoT,integratedBC&Edge}. Edge computing can be implemented into Blockchain to tap into edge computing's edge processing capabilities for the public architecture. Edge processing would be able to offer nodes connected at the "edge" of the network to have the same computation speed as nodes closer to the core network. But tests of edge processing in Blockchain has only been tested in permissioned type Blockchain \citep{blockchainIoTEdge}. Tests were only done for permission Blockchain, thus prompting the need to investigate permissionless Blockchain. 
     
     The ability to pool resources from public architectures would enable edge computing to work effectively with technologies centered around the network and architecture. This brings Software Defined Network (SDN) and Network Function Virtualization (NFV) \citep{futureEdgeForIoT} be relevant for discussion as part of the future technology that needs to be considered.

\subsection{Fog Computing}
    
    Fog Computing is described by \citep{Fog} as a system-level architecture distributing services and resources of computing, control, storage and networking anywhere with the continuum from Cloud to Edge. In this architecture, communication devices, like switches and routers, are able to provide various communication and computation features as their computational and storage resources are extended. The control, computing, data, security, and networking levels will allow for a robust standardization, unification, and convergence under the influence of this computing paradigm. This would give efficiency when implemented into Blockchain, where it will cut the necessary storage for network communication and transaction for both IoT and Blockchain. Fog computing's design is based on removing the distance and performance needed for network traffic. But fog computing's intentions are driven by marketing with user's interaction via advertising, entertainment, and Big Data analytical applications \citep{futureEdgeForIoT}. A notable flaw with fog computing is its fault tolerance level has not been extensively researched. The only results with current research on fault tolerance level are node failures with fog computing \citep{litReviewFog}. This flaw is resolved by implementing fog computing into Blockchain, by partitioning fog node clusters with fog nodes within a Blockchain \citep{coopBCFog}. This forms a Blockchain-based fog node cluster that uses a consensus algorithm to work with any computers in the network. This implementation also provides an increased level of machine-to-human communication, which is beneficial for IoT \citep{fogEnablerBCIIoT}. There is also network storage cost to consider, as fog computing would need to account for Big Data. Big Data would result in bottleneck performance for the network, which would affect the performance of fog computing and cloud computing. 
    
    There are two solutions to solving this bottleneck performance. The first solution is to have a federated learning Blockchain to assist fog computing \citep{decentralisedPrivacyFog}. This solution provides increased security and efficiency for the Blockchain, which would be suitable for decentralized privacy protection. The alternate proposal is fog computing being implemented with a novel "Plasma" framework Blockchain \citep{fogPlasmaFramework}. This "Plasma" framework enables fog nodes in the Blockchain to allow IoT to connect into the Blockchain. This solution solves the bottleneck by removing the need for large overhead storage or computation power for network transmission.
    
\subsection{Peer-to-Peer (P2P)}
    
    P2P technology is prevalent with Blockchain due to its association in the distributed network. P2P was popular in the age of privacy where users used P2P for file sharing. This provided a platform of anonymity which symbolized complete freedom on the Internet, which opened up an entire issue of digital piracy and DMCA. P2P is described as a peer being able to share resources with other peers in the network while maintaining equal roles and privileges within the network. P2P has an association with IoT for enabling both anonymity and decentralization at the cost of storage issues \citep{p2pPrivacy&Decentralization}. But in recent years, \citep{doNotDecentralize} states that there is a clear decline of pure P2P applications and software within the past years. But according to \citep{doNotDecentralize}, Blockchain with permissioned consensus may have the key to revitalize the decentralization and provide increased trust and dependability. 
    
\subsection{Data Networking}

    Despite the focus on having decentralization where every user is equal and not adhere to a central figure, network configuration plays an important role for standardization. Without this standardization, a multitude of issues may arise from issues of performance hindrance due to conflicting protocols, increased cost to accommodate different configurations, and reduced scalability and reliability due to conflicting configurations. Which brings in a difficult position of requiring an authoritative management to ensure both management and standardization of the network. The network configurations are maintained with network management applications. Network management applications have many approaches for handling networks, and each approach provides a different set of administrative and performance advantages. In a traditional network scheme, the Internet would operate similarly to a core network and allow computers to participate on the Internet via ISP's network infrastructure and data centers. But this scheme is avoided in the industry, due to its need for high expenditure of new equipment, accounting for clunky inherent configurations, and maintenance of the infrastructure. Therefore, a dynamic, scalable, and cheaper alternative is needed for maintaining the network.

     \begin{figure*}
         \centering
         \includegraphics[width=140mm]{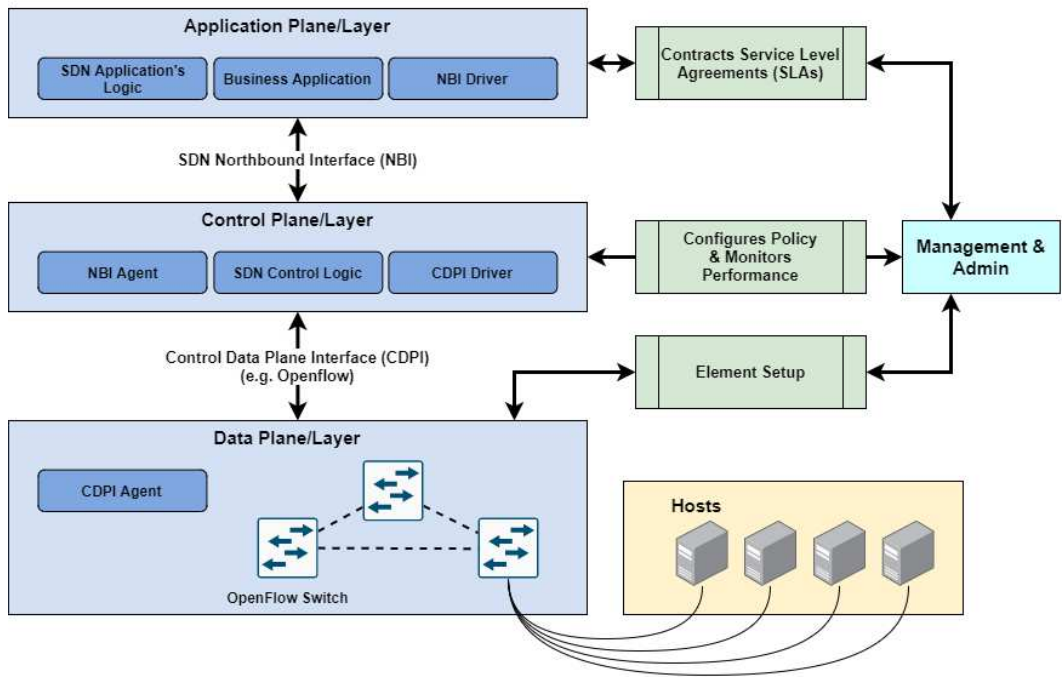}
         \caption{SDN Architecture}
         \label{fig:SDNArchitect}
     \end{figure*}

          \begin{figure*}
         \centering
         \includegraphics[width=140mm]{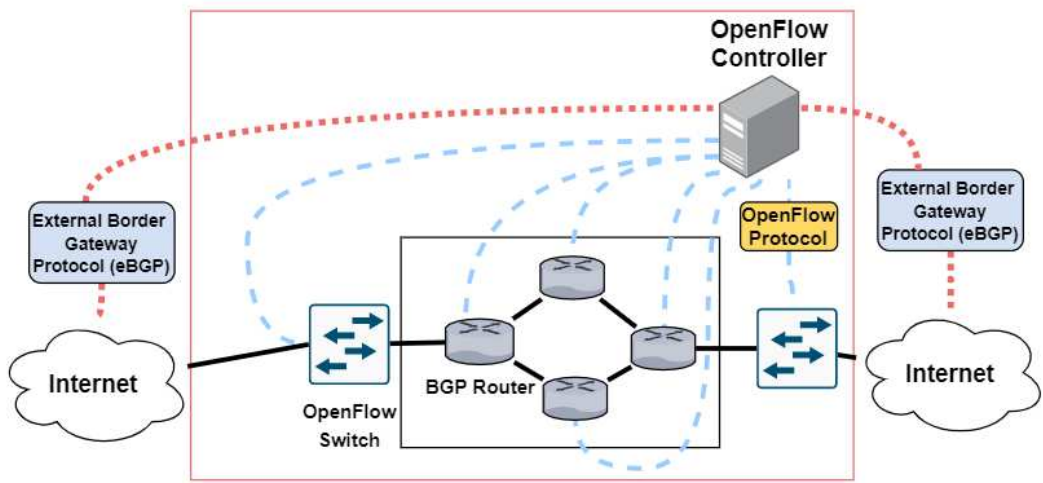}
         \caption{SDN Routing}
         \label{fig:SDNRoute}
     \end{figure*}
    
    \subsubsection{Software-Defined Networking (SDN)}
    
     These days, SDN has been loosely used by the networking industry for defining any network architecture that is operated by software. The original definition of SDN involves four components \citep{SDNsurvey}:
     
     \begin{enumerate}
         \item The ability to remove the control functionality for network devices.
         
         \item Usage of OpenFlow protocol, for its flow-based forwarding decision. This protocol is used to direct and manage network traffic between routers, switches, and vendors.
         
         \item An external controller which is a software platform that facilitates the control functionality while acting as a virtualization and resource vendor
         
         \item The ability of programming software application to operate on top of the controller and interact with underlying data plane devices
     \end{enumerate}
         
     The main attraction with SDN is its programmable feature to allow customizability to configure the network. SDN provides a dynamic configuration that operates from a central controller to be more efficient and customizable from traditional network infrastructure. When fog computing is implemented with an SDN-enabled Blockchain by deploying fog services, results have shown that there is an increase of performance and security for offloading data to the cloud while being cost-efficient \citep{sdnFogBC}. This implementation would result in a distributed Blockchain that uses SDN's controller to enable fog computing to offer low cost, secure, and on-demand access to edge nodes in the Blockchain. This proposed system would be scalable and secure enough to accommodate the expansion of IoT and the volume of data on the Internet and enable on-demand for low latency IoT devices. SDN's Architecture and routing can be seen in Fig.~\ref{fig:SDNArchitect} and Fig.~\ref{fig:SDNRoute}. 
    
    \subsubsection{Information-Centric Networking (ICN)}
    
    Information-Centric Network (ICN) is an alternate approach that centers around content data that is suited to the interest of the network \citep{ICNsurveyRDB}. ICN provides a cost-efficient and scalable method of handling the global expansion of IP traffic with its secure design of persistence and unique naming scheme for the data information. ICN consists of three components that make up the approach. 
    
    The first component is the Named Data Object (NDO), which is a self-certifying name method that is applied to an information data's metadata to give a unique identity. NDO consists of a unique identifier, the data, and the metadata \citep{ICNreview}. NDO adopts two types of naming schemes, which both offer unique names and security for the NDO. The first naming scheme is a hierarchical scheme that provides an aggregated approach for prefixes of the NDO. The second naming scheme is a self-certifying scheme which is done by embedding a hash containing the prefixes into data \citep{ICNsurvey}. 
    
    The second component is the Naming and Security of the information data, which encompasses the concept of establishing the identity of independent information data that is outside the network \citep{ICNsurvey}. This component consists of two schemes as well. The first scheme being Name Resolution Service (NRS) where it uses an external entity to interpret the name of the NDO after mapping the named data. But NRS suffers from a single point of failure due to the funneling of information data to an external entity for interpretation. The second scheme revolves around direct routing from the data requester to the data source of the network. This is heavily dependent on algorithms to find the properties needed to identify the namespace for both the requester and the data source. 
    
    The third component is the Application Programming Interface (API), which is used to request and deliver NDO around the network \citep{ICNsurvey}. The node that provisions the NDO is called a source/producer, who controls the publishing of the NDO in the network. NDO is requested by client/consumer through calling the NDO's name, which is done by request, finding, subscribing, or setting one of NDO's metadata as an interest. There are many approaches for managing how NDO is requested, from PSIRP where it is built on a subscription-based approach or CURLING where it supports location parameters.
    
    The fourth component is caching, where it is used to satisfy NDO requests by allowing nodes to hold a copy of the NDO in its cache. This application of caching allows ICN to apply edge computing and P2P for an in-network edge of "transparent web cache". Although the caching is simple by design, this can be improved with edge caching. As simulations have shown that current caching can be improved with edge caching to accommodate IoT for increased efficiency of data distribution for the ICN \citep{ICNreview}.

     \begin{figure*}
         \centering
         \includegraphics[width=140mm]{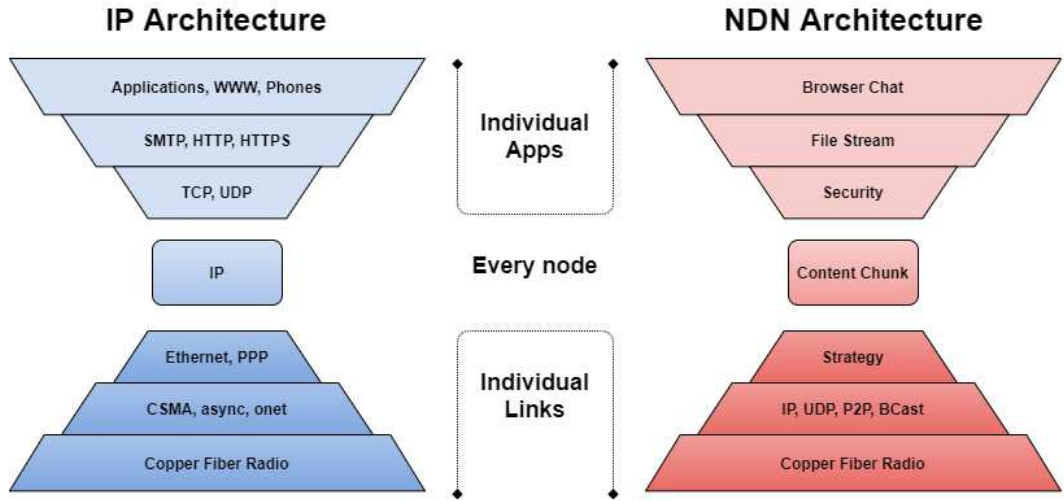}
         \caption{The IP and NDN Architecture Hourglass}
         \label{fig:IP&NDNhourglass}
     \end{figure*}
     
    \subsubsection{Content-Centric-Networking (CCN) \& Named-Data Networking (NDN)}
    
    The CCN is an architecture that is part of the ICN that centers around making content nameable and routable within the network. CCN communicates in the network through named data, as opposed to TCP/IP's approach of using IP addresses \citep{CCNinIoT}. CCN is able to improve the existing method of routing and forwarding from TCP/IP due to the named data. This improvement is achieved by computers fetching data with appropriately labeled names. This is later improved with Named Data Networking (NDN).
    
    NDN is an evolution of CCN where it uses the same approach of communication as CCN with named data \citep{introNDN}. NDN is designed to take advantage of rising new technology to meet the onset of demands such as Big Data that would make TCP/IP obsolete \citep{introNDN}. NDN's vision is to reshape the TCP/IP's hourglass structure by replacing IP with Content Chunks that are named data \citep{surveyNDNCCNinIoT}, which can be seen in Fig. ~\ref{fig:IP&NDNhourglass}. NDN can combine the networking aspect, storage expansion for the onset of Big Data, and the process of fetching data into a unified system to match and even overcome TCP/IP on meeting IoT's challenges on the network layer \citep{NDNforIoT}. NDN would give IoT a scalable, secure, energy-efficient, and heterogeneous system due to its functionalities. This benefit for IoT is also further reinstated with the proposal of introducing Fog Computing with NDN \citep{ICNandFOG}, where a smarter and more efficient approach in storage and resource provisioning to increase the performance of data transmission, caching, and improved security on the NDN. 
    
    In our opinion, NDN draws parallel to how Semantic technology is applied to the TCP/IP architecture in its concept of naming data. This parallel makes NDN capable of accomplishing the melding between Semantic technology and TCP/IP architecture with naming data chunks on the Internet by providing links, relevancy, and meaning to the data chunks.
    
\section{Discussion} \label{sectionDiscussion}
With new ideas and iterations of systems being discussed for development in the current tech industry, there would be point whereby decisions are needed to be made for dictating the directions of how Blockchain technology is utilized in the tech industry. The first topic discusses about trade-offs between technology, in our opinion the trade-off can be considered as an ongoing discussion of adopting new technologies to replace and improve legacy technologies operating within a system. It is through these trade-offs that forms new standardization for the future industries, and the decision to adopt Blockchain technology as a new norm is something to consider. The second topic discusses the relationship between the Internet and the impact of decentralization, where it discusses why we should decentralize and not allow monopolization from ISP. The third topic discusses development trends that are seen currently within the tech industry, as centralization from IoT and development of quantum computing poses a unique situtaion for the future development of the Internet. The fourth topic revolve around recentralization from the Internet, where it discusses the possibility of centralizing from the decentralization within Web 3.0. The fifth topic discusses battlefield implementation with IoT, where gathering and utilization of battlefield information through current and future technology such as Graphchain and NDN would enable the next step of cyber warfare. The last topic discusses the Merkle Tree where it is used the encryption of information to be stored in a ledger. As the Merkle tree is the only hash-based data structure used in Blockchain, would there be other alternatives to replace Merkle Tree like the proposed Verkle Tree\citep{KuszmaulVerkleTree}?

\subsection{Trade-off between technologies}
Trade-offs are always a concern when implementing new technologies to replace a new architecture, that is why there are different proposals for achieving Web 3.0. In our case, Blockchain has trade-offs occurring with the future technologies that we have proposed in Section~\ref{sectionFutureTech}. There are two notable trade-offs that need to be decided from this paper. \\
Graphchain is considered as an upgrade version of Blockchain in terms of optimization. However, there is a trade-off with Graphchain, where the optimized routes will eventually be centralized due to the route taken with common descendants. This makes the decision to decide how centralizability should a Blockchain have for Internet architecture. There is also the case of cloud computing's standardization, where there is a risk of reduced performance and scalability if we use middlewares for communication standardization. Deciding which technology to implement would be a challenge on itself, as balancing the trade-offs between technologies would be a hurdle in the advancement of developing the internet architecture.

\subsection{Relationship between the Internet and decentralized infrastructure}
The push for a decentralized architecture has resided within the Internet community, only to be reinforced with the incident from Net Neutrality. ISPs have complete control over the user's Internet with its monopolization of network flow for users connecting through the Internet, which was further discussed in Section~\ref{sectionChallengesCentral}. This monopolization from the ISP allows exploitation and abuse from large corporation. With how much personal information being linked due to social media's influence, it's no surprise that it is easy to trace a user's personal information based on techniques like social engineering. A decentralized architecture is the proposed solution, where its anonymity is used to prevent misuse of personal information. This leads to the outcry of having a decentralized architecture to distinguish users away from needing a centralized node, despite the drawbacks came from the initial first generation of Blockchain.

\subsection{Development Trends}
The current trend of the Internet is driven by the impending arrival of IoT. As the days of bulky computers are gone, comes the influx of new smart devices that would interact with the Internet architecture. Now the question lies on how the IoT interacts with the proposed Blockchain Internet architecture. One trend that is consistently shown in news outlets is smart devices being linked with each other in a network to form a high-tech lifestyle where smart devices connected in the network can be operable with a single smart device. Another future trend is quantum computing, as it brings optimization features for the future decentralized Internet with its quantum communication. Quantum communication would be able to outperform the limits of traditional sender-receiver communications. This communication is done by entangling quantum nodes to multiple levels of entanglement, which results in a heterogeneous multi-level entanglement network structure as noted by \citep{quantumDecentralised}. This network structure would then result in an efficient decentralized routing, which would be beneficial with the onset of exponential growth in information in the future Internet.

\subsection{Re-centralization}
Although the goal of this paper is to achieve decentralization for a future internet architecture, it brings up the question of how the Internet developed into its current centralized state. Web 1.0 was designed to be decentralized, only to be centralized in Web 2.0. This migration to Web 2.0 brought new centralized services that allowed the Web to have more functions and be more optimized than Web 1.0. The real challenge comes during the implementation of the decentralized Web 3.0 or dWeb. Would it be possible to offer the same optimization and efficiency of the centralized services, but in a decentralized way? A major aspect to consider is the personal data of a user, where a centralized architecture would provide a higher quality of life in personalization of applications and advertisements based on personalized information and profiling of users. But in the event of removing this feature to allow complete decentralization of the architecture to be an acceptable loss? This trade-off of quality of service would occur at the migration towards decentralization. Unless a new design of architecture that can preserve the services while maintaining a decentralized Internet is proposed, this would remain a huge issue. This is a huge conundrum in itself with personal data, as a decentralized architecture would present a situation where nobody could be held accountable for events that occurs. This brings the discussion of the practicality of data centers, with current investors steering towards the idea of investing in bigger data centers to account for the exponential growth of information on the Internet. However, with a Blockchain-enabled decentralized architecture, it would be possible to implement the services of data centres into individual nodes of the Internet, ensuring a probable solution that is cheaper, scalable, and efficient. But diving into a purely decentralized Internet would not be an ideal setting in the current world's reliance on a centralized governing figure such as the government and financial banks. This is caused by the concerns disruption of balance in their respective industry due to no governing forces as any updates are done via majority voting without a supervision forces. A balance is needed to provide for both centralized and decentralized in this aspect, as a purely decentralized network without a governing figure would ensure possible chaos without supervision.

\subsection{Battlefield implementation with IoT}
In the onset of a decentralized infrastructure, a unique situation comes from the attempted implementation of IoT into future battlefield situation with relevant network technology such as Graphchain and NDN \citep{IoTGraphBF,IoTNDNBC-BF}. By incorporating information data about battlefield information such as ammunition, troops, and enemy intelligence, this would change how current warfare is engaged. Information plays a vital role in the battlefield, as it provides benefits on how a commander would able to make quick and decisive tactical decisions based on on-site real-time information. Incorporating military aspects into a decentralized network infrastructure with Blockchain implementation seems to be a possible future. Integration of cyber warfare is already in the present, so it would be the next step of information warfare.

\subsection{Merkle Tree}
The Merkle tree is an important hash-based data structure used for optimized distribution and verification of the hashed ledger in Blockchain. This data structure allows each node to optimize the storage of multiple ledgers. This Merkle Tree is also used in the projects that are mentioned in Section \ref{types-of-decentralisation} for encoding files to be distributed around the decentralized network. 

The Merkle tree is used to encrypt multiple information many times to reach an eventual Merkle root, which houses multiple information of a single ledger. This Merkle root is then used to verify the integrity of the ledger for every decryption that has been executed on the ledger, to verify the hash's information. This brings the question of Merkle Tree being the only option for encryption in Blockchain, and if there are any other alternatives or optimized encryptions that can be considered. Although the Merkle tree can be re-purposed into a file system \citep{merkleFileSystem} where it is a decentralized network of P2P and is capable of being expanded. This brings back the argument of relying on Merkle Tree. Alternative encryption has been proposed with Verkle Tree by \citep{KuszmaulVerkleTree}, where it can optimize and reduce the bandwidth needed for consensus protocols to communicate in the network. However, Merkle Tree is still in its testing phases with limited resources and results shown, therefore leading us back to the Merkle Tree. This brings us back the question of would there be an alternative in encryption of the ledger that is better than the Merkle Tree.

\section{Conclusion} \label{sectionConclusion} 

This paper delved into the recommendation of Blockchain and how it is an effective enabler in achieving a decentralized Internet. Although there are other methods of achieving decentralization, we are confident with the choice of using Blockchain as an enabler to decentralize the Internet. From this paper, we understood that the current Internet architecture suffers from a myriad of issues as discussed in Section \ref{sectionChallengesCentral}, and proposed that using Blockchain would solve those issues. It is also discovered that the consensus algorithm would play a vital role in determining the level of power a node holds within the network, and how the network should communicate. From the list of consensus algorithms that have been discussed in Section \ref{sectionConsensus}, three algorithms which are Proof-Of-Property, Paxos, and Proof-Of-Authority, stood out as options for handling the nodes in the Blockchain. With upcoming technologies being constantly introduced into the industry, there would be better and more optimized technologies that can replace the proposed technologies that have been proposed in this paper. 

From this study, we have identified and investigated two important Blockchain research aspects which have key roles in feasibility of achieving a decentralized Internet using Blockchain. First, being the consensus algorithms that provide the needed decentralization but in factors of different optimization and achieving consensus. Second, the relevant technology which would reduce the flaws of Blockchain and help Blockchain to succeed in decentralizing the Internet. The survey that this paper has provided on Blockchain will help in providing coordination in achieving decentralization for the Internet.


\bibliographystyle{unsrt}
\bibliography{main.bib}

\break

\break

\end{document}